\documentclass[12pt,draftclsnofoot,onecolumn]{IEEEtran}
\usepackage{cite}
\ifCLASSINFOpdf
  \usepackage[pdftex]{graphicx}
\else
  \usepackage[dvips]{graphicx}
\fi
\usepackage[cmex10]{amsmath}
\usepackage{array}
\usepackage{theorem}
\usepackage[caption=false,font=footnotesize]{subfig}
\usepackage{fixltx2e}
\usepackage{url}

{\theorembodyfont{\rmfamily}
   }
{\theorembodyfont{\rmfamily}
   }
{\theorembodyfont{\rmfamily}
   }
{\theorembodyfont{\rmfamily}
   }

\begin{document}

\title{Coding with Scrambling, Concatenation, and HARQ for the AWGN Wire-Tap Channel: A Security Gap Analysis}

\author{\IEEEauthorblockN{Marco Baldi,~\IEEEmembership{Member,~IEEE,} Marco Bianchi, and Franco Chiaraluce,~\IEEEmembership{Member,~IEEE}}
\thanks{Copyright (c) 2010 IEEE. Personal use of this material is permitted. 
However, permission to use this material for any other purposes must 
be obtained from the IEEE by sending a request to pubs-permissions@ieee.org.

Part of the material in this paper has been presented 
at the IEEE Information Theory Workshop (ITW 2010),
Dublin, Ireland, August 30--September 3, 2010
and at the IEEE ICC 2011 Workshop on 
Physical Layer Security, Kyoto, Japan, June 5, 2011.

M. Baldi, M. Bianchi and F. Chiaraluce are with Dipartimento di Ingegneria dell'Informazione,
Universit\`a Politecnica delle Marche, Ancona, Italy
(e-mail: \{m.baldi, m.bianchi, f.chiaraluce\}@univpm.it).

This work was partially supported by the MIUR project ``ESCAPADE'' (grant no. RBFR105NLC) under the ``FIRB - Futuro in Ricerca 2010'' funding program.

}}

\maketitle

\begin{abstract}
This study examines the use of nonsystematic channel codes to obtain
secure transmissions over the additive white Gaussian noise (AWGN) wire-tap channel.
Unlike the previous approaches,
we propose to implement nonsystematic coded transmission by
scrambling the information bits, and characterize the bit error rate of
scrambled transmissions through theoretical arguments and numerical simulations.
We have focused on some examples of Bose-Chaudhuri-Hocquenghem (BCH) and 
low-density parity-check (LDPC) codes to estimate the security gap,
which we have used as a measure of physical layer security, in addition to the bit error rate.
Based on a number of numerical examples, we found that such a transmission technique can outperform
alternative solutions. In fact, when an eavesdropper (Eve) has a worse channel than the
authorized user (Bob), the security gap required to reach a given level of security
is very small.
The amount of degradation of Eve's channel with respect to Bob's that is needed to achieve
sufficient security can be further reduced by implementing scrambling and descrambling 
operations on blocks of frames, rather than on single frames.
While Eve's channel has a quality equal to or better than
that of Bob's channel, we have shown that the use of a hybrid automatic repeat-request (HARQ) protocol with 
authentication still allows achieving a sufficient level of security.
Finally, the secrecy performance of some practical schemes has also been measured
in terms of the equivocation rate about the message at the eavesdropper and compared 
with that of ideal codes.
\end{abstract}


\section{Introduction}
Transmission security is often implemented at protocol layers higher than the physical
one, by exploiting cryptographic techniques based on computation assumptions.
These schemes rely on the existence of one or more cryptographic keys that
must be known by legitimate users and protected from eavesdroppers.
On the contrary, when security is implemented at the physical layer, all receivers
are perfectly aware of the encoding and transmission procedures,
without the need of any shared secret.
In this case, security is only based
on the differences between the channels experienced by authorized and unauthorized users.
On the other hand, exploiting these asymmetries often requires knowledge of the channel, while 
this assumption is not required in traditional cryptography.
Therefore, physical layer security can be viewed as a substrate helping to reduce the
complexity of cryptographic techniques at higher layers. 

Starting from these premises, it is important to investigate which transmission techniques can
be used for physical layer security and which of them are able to
exploit even small differences between the channels of the authorized and unauthorized
users.

\begin{figure}[htb]
\begin{centering}
\includegraphics[width=90mm,keepaspectratio]{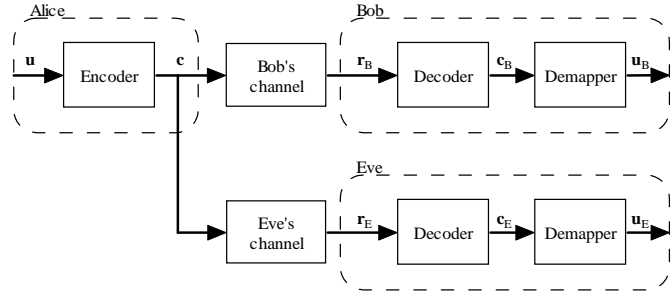}
\caption{Block scheme of a wire-tap channel. \label{fig:WireTap}}
\par\end{centering}
\end{figure}

For this purpose, we consider the well-known \textit{wire-tap channel} model \cite{Wyner1975}, shown
in Fig. \ref{fig:WireTap}, where a transmitter (Alice) encodes a message vector ($\mathbf{u}$)
into a codeword vector ($\mathbf{c}$) before transmitting it.
Alice's transmission is received by a legitimate receiver (Bob) and an eavesdropper (Eve),
and the channel that separates Alice from Bob is generally different from that between Alice
and Eve. Therefore, the vector received by Bob ($\mathbf{r}_\mathrm{B}$) is 
different from that gathered by Eve ($\mathbf{r}_\mathrm{E}$).
It follows that the two codeword vectors ($\mathbf{c}_\mathrm{B}$ and $\mathbf{c}_\mathrm{E}$) that Bob and Eve 
obtain after decoding can also be different.
If this occurs, after inverting the encoding map, Bob and Eve obtain two estimates of the message vector,
noted by $\mathbf{u}_\mathrm{B}$ and $\mathbf{u}_\mathrm{E}$, with $\mathbf{u}_\mathrm{B} \neq \mathbf{u}_\mathrm{E}$, which
is the basis for physical layer security.
However, when Eve's probability of error is small, she might be able
to correct all errors in her observed codeword by exploiting her complete knowledge of 
the transmission technique, which is enough to break secrecy.
Therefore, it is important to avoid that this occurs by finding suitable transmission techniques.

Under an information theoretic viewpoint,
the wire-tap channel can be described through the secrecy capacity, defined as the
highest transmission rate at which Bob can achieve arbitrarily small error probability,
while the mutual information between $\mathbf{u}$ and $\mathbf{r}_\mathrm{E}$ goes to zero as the blocklength goes to infinity.
To take into account the effect of encoding, the equivocation rate at the eavesdropper,
defined as the conditional entropy of the secret message, given Eve's observation,
can also be used as a measure of secrecy.
Despite the fact that both the secrecy capacity and equivocation rate provide important estimates 
of secrecy, when we adopt practical coding and modulation schemes, another valuable
metric of security is the bit error rate (BER) \cite{Klinc2011}.
In fact, when Eve experiences a BER of about $0.5$ and the errors are randomly
distributed, she is not able to get enough information 
on the transmitted message. Ensuring values of Eve's BER close to $0.5$ is also useful 
for cryptographic techniques that work at higher layers and aim at reaching a 
prefixed level of computational security.

The BER is also used as a security measure in estimating the \textit{security gap}, which was first introduced by Klinc et al.
\cite{Klinc2009a}, and is defined as the quality ratio between Bob's and Eve's channels required
to achieve a sufficient level of physical layer security, while ensuring that Bob reliably receives
the information.
We suppose that Bob's and Eve's channels are corrupted by additive white Gaussian noise (AWGN)
with different signal-to-noise ratio (SNR): $\left. \frac{E_b}{N_0} \right|_\mathrm{B}$ is
Bob's channel energy per bit to noise power spectral density ratio, whereas
$\left. \frac{E_b}{N_0} \right|_\mathrm{E}$ is the same parameter for Eve's channel.
Similarly, $\left. P_e \right|_\mathrm{B}$ is Bob's bit error probability and $\left. P_e \right|_\mathrm{E}$ is that of Eve's.
Following the approach in \cite{Klinc2009, Klinc2009a, Klinc2011}, we aim at achieving 
a given performance in terms of reliability and security, that is,
$\left. P_e \right|_\mathrm{B} \leq \overline{\left. P_e \right|_\mathrm{B}}$
and $0.5 \geq \left. P_e \right|_\mathrm{E} \geq \overline{\left. P_e \right|_\mathrm{E}}$,
where $\overline{\left. P_e \right|_\mathrm{B}}$ and $\overline{\left. P_e \right|_\mathrm{E}}$
are suitably fixed thresholds.
Under an information theoretical viewpoint, achieving security does not
require Eve's error probability to be fixed. However, a high error probability 
for Eve is a desirable feature for practical secure schemes.

Starting from the curve of bit error probability as a function, $f$, of the SNR,
$\overline{\left. P_e \right|_\mathrm{B}}$ and $\overline{\left. P_e \right|_\mathrm{E}}$
can be expressed in terms of $\frac{E_b}{N_0}$ as follows:
\begin{equation}
\left\{ 
\begin{array}{l}
\overline{\left. P_e \right|_\mathrm{B}} = f\left(\left. \overline{\frac{E_b}{N_0}} \right|_\mathrm{B}\right), \\
\overline{\left. P_e \right|_\mathrm{E}} = f\left(\left. \overline{\frac{E_b}{N_0}} \right|_\mathrm{E}\right),
\end{array} 
\right.
\end{equation}
and the security gap is: 
\begin{equation}
S_g = \frac{\left. \overline{\frac{E_b}{N_0}} \right|_\mathrm{B}}{{\left. \overline{\frac{E_b}{N_0}} \right|_\mathrm{E}}}.
\end{equation}

According to its definition, it is important to keep the security gap as small as possible, so that the
desired security level is achieved even with a small degradation of Eve's channel with respect to that of Bob's.
It is evident that the security gap depends on the steepness
of the curve $P_e = f\left(\frac{E_b}{N_0}\right)$: The steeper the slope is, the smaller will be the security gap.

Some previous works have been devoted to examine the transmission techniques that are able
to reduce the security gap.
A first requirement is to avoid
systematic transmission, which would directly expose the secret information bits.
In \cite{Klinc2009, Klinc2009a, Klinc2011, WongWong2011a, WongWong2011}, this target has been achieved
through punctured codes, by associating the secret information bits to punctured bits.
Thus, they are no longer transmitted over the channel and must be recovered by the receiver
from the nonpunctured part of the codeword.
In those papers, the authors considered punctured low-density parity-check (LDPC) codes and proved that 
for a fixed rate, puncturing is able to guarantee a significant reduction in the security gap 
with respect to nonpunctured (systematic) transmission.
The analysis in \cite{Klinc2009, Klinc2009a, Klinc2011} assumed that Eve's channel is more noisy than Bob's;
that is, a security gap greater than $1$.
When the security gap is $1$, that is, Eve's channel has the same quality as that of Bob's,
a sufficient level of security can still be achieved by introducing a hybrid automatic repeat-request (HARQ) protocol 
with authentication \cite{Tang2009}.

Although being effective in reducing the security gap, punctured codes result in higher 
power consumption with respect to nonpunctured transmission \cite{Klinc2011}.
This fact will also result from the comparison presented in Section \ref{sec:five}.
To overcome such limit, we have proposed an alternative solution exploiting nonsystematic
transmission based on scrambling of the information bits \cite{Baldi2010}. 
This technique is able to achieve a strong reduction in the security gap 
that becomes comparable with (and even better than) that obtained through puncturing.
Moreover, this result is achieved without increasing the transmitted power,
because the frame error rate is unaffected by scrambling.
We have also shown that the security gap can be further reduced by combining concatenation 
with scrambling \cite{Baldi2011}.
In this case, the information frames are grouped into blocks and scrambled/descrambled together,
thus increasing the effect of scrambling.

In the present study, we have merged and extended the earlier works presented in \cite{Baldi2010} and \cite{Baldi2011}, 
aiming at providing a thorough analysis of nonsystematic coded transmission for the wire-tap channel.
First, we have developed some theoretical tools that allow characterizing the BER performance of
scrambled transmissions. Based on these tools and numerical simulations, we have provided
some examples of the estimation of the security gap for the cases of hard-decoded classical block codes
such as Bose-Chaudhuri-Hocquenghem (BCH) codes, and modern soft-decoded block codes such as LDPC codes.
This allows comparing systematic and nonsystematic transmission techniques under both
reliability and security viewpoints, thus providing useful insights on the design of
coded transmission schemes for these purposes.
We have also considered the combined use of coding with scrambling, concatenation, and HARQ.
We have fixed a set of code parameters and provided a wide set of performance assessments,
thus extending the comparison among these techniques.
In addition, we have estimated the complexity of systematic and nonsystematic transmission 
techniques, allowing evaluation of the tradeoff between performance and complexity.
We have also extended the analysis to the case in which Eve's channel has a better quality than
that of Bob's, and we have shown that by using HARQ in conjunction with concatenated scrambling,
a sufficient level of security can still be achieved.
Finally, we have introduced the equivocation rate as a security measure in this context
to confirm that practical scrambled LDPC codes can outperform solutions based on puncturing 
and achieve secrecy performance very close to the theoretical bounds.

The rest of the paper is organized as follows. 
In Section \ref{sec:two}, we have introduced the notation and described the system that exploits coding with scrambling;
in Section \ref{sec:three}, we have studied the effect of scrambled transmission with different families of
linear block codes;
in Section \ref{sec:arq}, we have investigated the effect of HARQ protocols on coded transmission with scrambling;
in Section \ref{sec:Secrecy}, the equivocation rates achievable by scrambled codes have been evaluated;
and finally, in Section \ref{sec:six}, conclusion is presented.

\section{Notation and Related Work}
\label{sec:two}

In the considered AWGN wire-tap channel model, Alice sends a secret message 
in the form of a $1 \times k_s$ binary vector $\mathbf{u}$.
Before transmission, the secret message is encoded by Alice into a $1 \times n$ 
binary codeword $\mathbf{c}$, with $n \geq k_s$, which is then transmitted over the 
channel.
The \textit{secrecy rate} $R_s$ is defined as the ratio between the secret message
length and the transmitted codeword length; that is, $R_s = k_s/n$.
In fact, Alice could use a code with dimension $k \geq k_s$
and rate $R = k/n \geq R_s$ by padding her secret message with $k - k_s$ nonsecret information bits,
before encoding it.
In the following, however, we will limit to consider $k = k_s$, so that $R = R_s$.
This is because, as shown in \cite{Klinc2011}, choosing $R > R_s$ could help to reduce the security gap,
but also gives some loss in terms of SNR.
Therefore, as the need to save power is of primary importance in most of the applications, the choice $R = R_s$
is preferred. 

According to the wire-tap channel model shown in Fig. \ref{fig:WireTap}, we denote the codewords 
decoded by Bob and Eve as $\mathbf{c}_\mathrm{B}$ and $\mathbf{c}_\mathrm{E}$, respectively.
Some remarks concerning the decoding strategies are useful.
Under a cryptographic viewpoint, the assumption that an attacker uses the optimal decoder
is required to estimate the minimum security level of the system.
In the case of LDPC codes, this can be done through asymptotic analyses, possibly based on density evolution \cite{Richardson2001}, which accurately describes the performance of belief propagation decoders on arbitrarily long codes.
Moreover, the belief propagation decoder is asymptotically equal to the bitwise maximum a posteriori probability (MAP)
decoder. Thus, the results of density evolution
can be referred to the use of a bitwise MAP decoder.
Obviously, for finite-length codes, the use of belief propagation, though being the
best-known LDPC decoding algorithm with acceptable complexity, must be considered
as a suboptimal choice.

However, the use of suboptimal decoders is a further step in the research, especially with the
purpose of assessing the security and performance of practical transmission schemes.
Therefore, we make some precise assumptions on the decoder used by the eavesdropper.
When dealing with BCH codes, we suppose that Eve uses a bounded-distance hard-decision decoding
algorithm, such as the Berlekamp-Massey decoder \cite{Massey1969}.
When working with LDPC codes, we suppose that Eve uses the sum-product algorithm (SPA)
with log-likelihood ratios (LLR) \cite{Hagenauer1996}.
In Section \ref{sec:Secrecy}, we have estimated the performance of
LDPC codes through density evolution, thus modeling a bitwise MAP decoder.
Throughout the paper, we have considered that Bob uses the same decoding technique as Eve. 
In fact, it is not necessary to consider that Eve's
decoder is better than Bob's to make a comparison among different 
transmission techniques. 





In the scheme that we have considered, Alice implements encoding as follows:
\begin{equation}
\mathbf{c} = \mathbf{u \cdot S \cdot G},
\label{eq:Encoding}
\end{equation}
where $\mathbf{G}$ is the $k \times n$ generator matrix of an $(n, k)$-linear block code in systematic
form and $\mathbf{S}$ is a nonsingular $k \times k$ binary scrambling matrix.
Owing to its systematic form, $\mathbf{G}$ can also be written as 
$\mathbf{G} = \left[ \mathbf{I} | \mathbf{C} \right]$, where $\mathbf{I}$ is a $k \times k$
identity matrix and $\mathbf{C}$ is a $k \times \left(n-k\right)$ matrix representing
the parity-check constraints.
Thus, encoding simply consists of replacing
the information vector $\mathbf{u}$ with its scrambled version $\mathbf{u' = u \cdot S}$,
and then applying the linear block code given by $\mathbf{G}$.
According to the physical layer security principle, both $\mathbf{S}$ and $\mathbf{G}$
are made public and both of them are necessary for decoding: The knowledge of the code (through
$\mathbf{G}$ or, equivalently, the parity-check matrix $\mathbf{H}$) is necessary to exploit
its error correcting capability, while $\mathbf{S}$ (in fact, its inverse) must be used for
descrambling $\mathbf{u'}$ into $\mathbf{u}$. Therefore, 
both $\mathbf{G}$ and $\mathbf{S}$ are known to Bob and Eve, and the security of the system
does not rely on any secret information.

Based on the above-mentioned assumptions, the encoded codeword can also be written as 
$\mathbf{c} = \left[\mathbf{u \cdot S} | \mathbf{u \cdot S \cdot C}\right] = \left[\mathbf{c}_l | \mathbf{c}_r \right]$,
where $\mathbf{c}_l$ is the vector containing the first $k$ bits of $\mathbf{c}$, while $\mathbf{c}_r$
collects its last $r = n-k$ bits.
Obviously, both Bob's and Eve's channels introduce errors. However, as mentioned earlier, $\left. \frac{E_b}{N_0} \right|_\mathrm{B}$ should be large enough to ensure that Bob's decoder is able to correct all errors with very high probability, thus recovering $\mathbf{u}_\mathrm{B} = \mathbf{u} = \mathbf{c}_l \cdot \mathbf{S}^{-1}$. On the contrary, $\left. \frac{E_b}{N_0} \right|_\mathrm{E}$ should be small enough to ensure that the codeword obtained by Eve after decoding is still affected by an error vector $\mathbf{e} = \left[\mathbf{e}_l | \mathbf{e}_r \right]$.
In this case, Eve gets:
\begin{equation}
\mathbf{u}_\mathrm{E} = \left( \mathbf{c}_l + \mathbf{e}_l \right) \cdot \mathbf{S}^{-1} = \mathbf{u} + \mathbf{e}_l \cdot \mathbf{S}^{-1}.
\label{eq:ErrorSpreading}
\end{equation}
Therefore, owing to multiplication by $\mathbf{S}^{-1}$, descrambling can propagate the residual errors.

To estimate the performance achievable by scrambled transmissions, we
can refer to the ideal case that we denote as \textit{perfect scrambling}, in
which even one residual channel error produces maximum uncertainty.
In other terms, under the hypothesis of perfect scrambling, a single residual
bit error in vector $\mathbf{e}_l$ is sufficient to ensure that half of the information
bits are in error after descrambling and that the error positions are randomly distributed. 
Perfect scrambling is an expression of the \textit{strict avalanche effect}, which is one of the most desirable properties of cryptographic algorithms. Good avalanche features are important to ensure that an algorithm is not susceptible to statistical attacks and ensure randomness of the ciphertext \cite{Heys1995}.
In practice, perfect scrambling can be approached by using a matrix $\mathbf{S}^{-1}$ with a high density of ones.
The best scrambling effect is obtained when the density of 
$\mathbf{S}^{-1}$ is $0.5$, but
a lower density of $\mathbf{S}^{-1}$ can suffice to approach perfect scrambling, as shown next.

The propagation effect of scrambling matrices on residual errors can be further
increased by implementing the scrambling and descrambling operations on blocks
of frames, rather than on single frames.
Let us consider collecting $L$ consecutive information frames in a vector
$\overline{\mathbf{u}} = \left[\mathbf{u}_1 | \mathbf{u}_2 | \ldots | \mathbf{u}_L \right]$.
Scrambling can be directly applied on the whole $L$-frame block as:
\begin{equation}
\overline{\mathbf{u}'} = \overline{\mathbf{u}} \cdot \overline{\mathbf{S}},
\label{eq:ConcScramb}
\end{equation}
where $\overline{\mathbf{S}}$ is a scrambling matrix with size $kL \times kL$; thus, $\overline{\mathbf{u}'}$ is a scrambled version of the $L$-frame block $\overline{\mathbf{u}}$.
After scrambling, vector $\overline{\mathbf{u}'}$ is divided into $1 \times k$ subvectors
$\left[\mathbf{u}'_1 | \mathbf{u}'_2 | \ldots | \mathbf{u}'_L \right]$,
that are encoded and transmitted separately.
Both Bob and Eve must collect their received frames into blocks
of $L$ frames before applying the block descrambling matrix $\overline{\mathbf{S}}^{-1}$.
Therefore, after descrambling, Eve gets a block of $L$ information frames
$\overline{\mathbf{u}_\mathrm{E}} = \overline{\mathbf{u}} + \overline{\mathbf{e}_l} \cdot \overline{\mathbf{S}}^{-1}$,
where $\overline{\mathbf{e}_l}$ is the vector formed by the concatenation of
the error vectors affecting the $L$ information frames.

Implementing scrambling and descrambling on blocks of $L$ frames (with $L>1$)
increases the error propagation effect, because a single residual bit error
in any of the $L$ frames can be spread over all of them.
The concept of perfect scrambling can be extended as follows.
Under the perfect scrambling condition, a single bit error
in one of the $L$ frames ensures that after descrambling: i) all the $L$ 
frames are in error and ii) half of the bits in each frame are in error and
their positions are random.
The perfect scrambling condition can also be approached for concatenated scrambling
using practical system parameters, as will be shown in the following. 

We noticed that the increase in complexity due to concatenation is limited, because the code continues
 to work on single frames. 
The only effect is a latency increase that, however, can be taken under control through the choice of $L$ and $k$.

\section{Scrambled codes for the wire-tap channel}
\label{sec:three}

To assess the performance, we need an estimate of the bit error probability 
($P_e$) and frame error probability ($P_f$) for Bob and Eve with and without scrambling.
For such purpose, we first refer to an explicative case, denoted as ``unitary rate coding.''
In the absence of scrambling, we have a classical uncoded transmission (i.e., using a
fictitious code with $\mathbf{G} = \mathbf{I}_k$, the $k \times k$ identity
matrix).
With scrambling, \eqref{eq:Encoding} yields unitary rate encoding through
matrix $\mathbf{S}$.
After having analyzed the case of unitary rate coding, we will investigate the more realistic
scenario of hard-decoded \textit{t}-error correcting codes, such as BCH codes.
Finally, we will consider LDPC coding as an example of
state-of-the-art soft-based coding scheme, and compare the proposed approach with 
that based on puncturing.

\subsection{Unitary Rate Coding}
\label{subsec:UnitaryRate}

Under the assumption of using binary phase shift keying (BPSK) modulation, the bit and frame error
probabilities are given by:
\begin{equation}
\left\{ 
\begin{array}{l}
P_e = \frac{1}{2}\mathrm{erfc}\left(\sqrt{\frac{E_b}{N_0}}\right), \\
P_f = 1 - \left(1 - P_e\right)^k. \\
\end{array} 
\right.
\label{eq:UnitaryUnscrambled}
\end{equation}

\subsubsection{Scrambling of single frames}
\label{subsubsec:UnitaryRateSingleFramse}

When scrambling is performed on single frames,
assuming a perfect scrambling condition, the bit
error probability after descrambling equals half the frame error probability
expressed by \eqref{eq:UnitaryUnscrambled}; that is:
\begin{equation}
P_e^{PS} = \frac{1}{2}\left\{1 - \left[1 - \frac{1}{2}\mathrm{erfc}\left(\sqrt{\frac{E_b}{N_0}}\right)\right]^k\right\}. \\
\label{eq:UnitaryPerfectScrambling}
\end{equation}
As perfect scrambling ensures Eve's maximum uncertainty,
it can be used as a bound on the security performance.
However, for real scrambling matrices, the bit error probability
for a unitary rate coded transmission can be estimated as follows.

%

We denote the column weight of matrix $\mathbf{S}^{-1}$ by $w \leq k$,
which we consider as regular for simplicity.
$P_j$ denotes the probability that a received $k$-bit vector contains $j$ errors
before descrambling, whereas $P_{i|j}$ is the probability that exactly $i$
out of $j$ errors are selected by a weight $w$ column of $\mathbf{S}^{-1}$.
The bit error probability 
after descrambling can be expressed as:
\begin{equation}
P_e^{S} = \sum_{j=0}^k{P_j}\sum_{\begin{subarray}{c} i=1\\ i \ \mathrm{odd} \end{subarray}}^{\min\left(j,w\right)}{P_{i|j}},
\label{eq:PeS}
\end{equation}
with
\begin{equation}
\left\{ 
\begin{array}{l}
P_j = {k \choose j}P_e^j\left(1-P_e\right)^{k-j}, \\
P_{i|j} = \frac{{j \choose i}{{k-j} \choose {w-i}}}{{k \choose w}}.
\end{array} 
\right.
\label{eq:PjPij}
\end{equation}

\begin{figure}[htb]
\begin{centering}
\includegraphics[width=83mm,keepaspectratio]{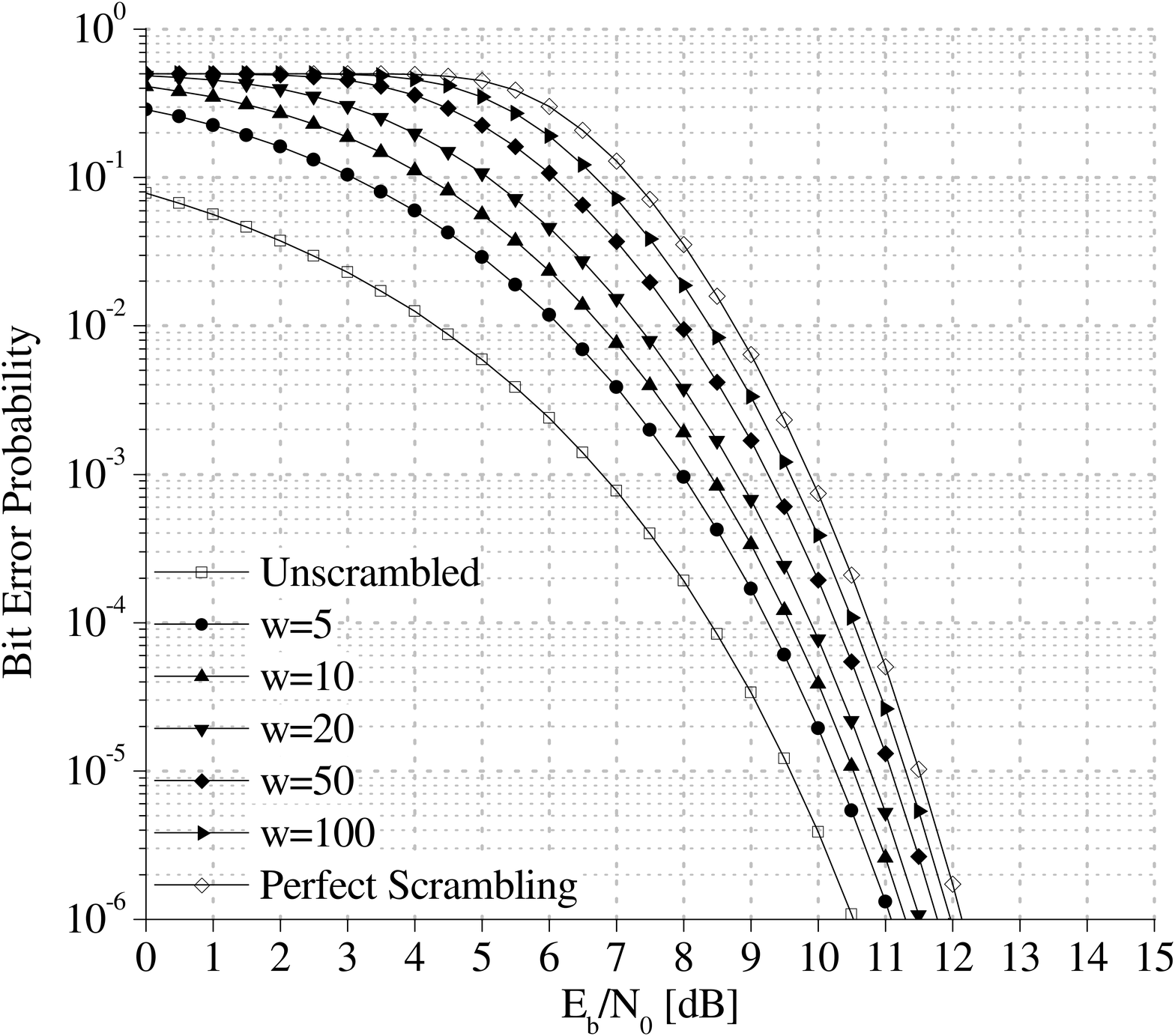}
\caption{Bit error probability with unitary rate coding ($k = n = 385$) in the absence of scrambling,
with perfect scrambling, and for different values of the descrambling matrix column weight ($w$).
\label{fig:UnitaryRate}}
\par\end{centering}
\end{figure}

As a numerical example, we consider $k = n = 385$ (this value of $k$ will also be of
interest subsequently) and compute the bit error probability for
several values of $w$.
Fig. \ref{fig:UnitaryRate} shows the dependence of $P_e^{S}$ on $w$ 
and the two limit cases of unscrambled transmission and perfect scrambling.

As we observe from the figure, the unscrambled transmission gives low values
of bit error probability even at a rather low SNR.
On the contrary, scrambling is effective in keeping the bit error
probability next to $0.5$ 
up to a rather high $\frac{E_b}{N_0}$ threshold (in the order of $4$ dB
with perfect scrambling).
In addition, scrambling helps to improve the slope of the 
$P_e$ curve, thus reducing the security gap.
As expected, perfect scrambling provides the highest error probability.
However, an $\mathbf{S}^{-1}$ matrix with density $\approx 0.26$ ($w = 100$) is able
to approach the performance of the perfect scrambler.
This confirms that it is not necessary to reach a matrix density equal to $0.5$
to approximate a perfect scrambler.
A lower density of $\mathbf{S}^{-1}$ is important under the complexity
viewpoint, because it is proportional to the number of operations needed by descrambling.

\subsubsection{Scrambling of concatenated frames}
\label{subsubsec:UnitaryRateConcatenation}

Let us consider the case in which scrambling is performed on $L$-frame blocks,
rather than on single frames.
In this case, the frame and bit error probabilities with perfect scrambling
can be estimated as:
\begin{equation}
\left\{
\begin{array}{l}
P_f^{L\textrm{-}PS} = 1 - \left(1 - P_f \right)^L, \\
P_e^{L\textrm{-}PS} = \frac{1}{2}P_f^{L\textrm{-}PS},
\end{array}
\right.
\label{eq:ConcPerfScrambUnitary}
\end{equation}
where $P_f$ is given by \eqref{eq:UnitaryUnscrambled}.
However, when real scrambling matrices are adopted,
the bit error probability can be
estimated starting from the bit error probability with
single frame descrambling, expressed by \eqref{eq:PeS}.

Let us consider a $kL \times kL$ block descrambling matrix
formed by $L \times L$ square blocks with size $k \times k$ and row/column
weight $w$.
In fact, such a descrambling matrix would be singular; hence,
in practice, it must have slightly irregular row and column weights.
However, under the error probability viewpoint, this yields negligible 
deviations from the case of regular matrices, considered here for the sake of simplicity.
After multiplication of each vector $\mathbf{e}_l$ by a $k \times k$ block, the probability that a bit is
in error is still $P_e^S$, given by \eqref{eq:PeS}.
For the block descrambling matrix that we have considered,
after block descrambling, each received bit can be seen as the 
sum of $L$ bits received after single frame descrambling. 
Therefore, its error probability can be estimated as:
\begin{equation}
P_e^{L\textrm{-}S} = \sum_{\begin{subarray}{c} i=1\\ i \ \mathrm{odd} \end{subarray}}^{L}{{L \choose i}\left(P_e^S\right)^i\left(1-P_e^S\right)^{L-i}}.
\label{eq:PeLS}
\end{equation}

\begin{figure}[htb]
\begin{centering}
\includegraphics[width=83mm,keepaspectratio]{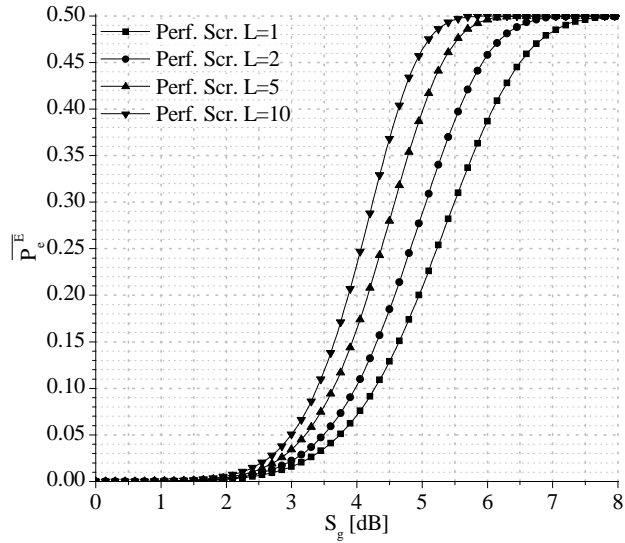}
\caption{Eve's bit error probability versus the security gap for 
$\overline{P_e^{\mathrm{B}}} = 10^{-5}$ with unitary rate coding 
($k = n = 385$) and different levels of concatenation ($L$),
under the hypothesis of perfect scrambling.
\label{fig:ConcPerfScrambUnitary}}
\par\end{centering}
\end{figure}

To evaluate the effect of concatenated scrambling on the security gap,
we again consider a unitary rate coded transmission with $k=n=385$ and fix 
$\overline{P_e^{\mathrm{B}}} = 10^{-5}$.
Fig. \ref{fig:ConcPerfScrambUnitary} reports the values of $\overline{P_e^{\mathrm{E}}}$, as a function 
of the security gap, for different values of $L$.
As we observe from the figure, the use of concatenated scrambling reduces the
security gap needed to reach high values of $\overline{P_e^{\mathrm{E}}}$ with respect to the case 
without concatenation ($L=1$).

Obviously, in the case of unitary rate coding, the performance of concatenated
scrambling over a block of $L$ frames, each with size $k$, is exactly coincident with that
of single frame scrambling with frame size $k' = kL$.
Thus, in this illustrative case, increasing the number of concatenated frames or the
frame size produces the same effect.
However, as it will be shown in the next section, this is no longer valid when error correcting codes are used.

\subsection{\textit{t}-Error Correcting Codes}

A linear block code with rate $R < 1$ can be used to increase the slope of the $P_e$ curves.
In this subsection, we consider an $(n, k)$ code
capable of correcting $t$ bit errors under hard-decision decoding, as it occurs for BCH codes.
As an example, in the following we will focus on the $(511, 385)$ BCH code, capable of correcting
$t = 14$ errors.
When this kind of codes is used, the frame and bit error probabilities at the receiver can be estimated as follows \cite{Torrieri1984}:
\begin{equation}
\left\{ 
\begin{array}{l}
P_f = \displaystyle\sum^n_{i=t+1} {n \choose i} P_0^i (1-P_0)^{n-i}, \\ 
P_e = \displaystyle\sum^n_{i=t+1}\frac{i}{n} {n \choose i} P_0^i (1-P_0)^{n-i}, \\ 
\end{array} 
\right.
\label{eq:tErrorCoding}
\end{equation}
where $P_0$ is the channel bit error probability, taking into account the bandwidth
expansion due to the code, that is:
\begin{equation}
P_0 = \frac{1}{2}\mathrm{erfc}\left(\sqrt{\frac{E_b}{N_0}\cdot\frac{k}{n}}\right).
\label{eq:P0}
\end{equation}

\subsubsection{Scrambling of single frames}
\label{subsubsec:tErrorCorrectingSingleFramse}

Let us first consider the case of single frame scrambling.
Starting from \eqref{eq:tErrorCoding}, the bit error probability 
with perfect scrambling is obtained as:
\begin{equation}
P_e^{PS} = \frac{1}{2}P_f = \frac{1}{2} \displaystyle\sum^n_{i=t+1} {n \choose i} P_0^i (1-P_0)^{n-i}.
\label{eq:tErrorCodingPS}
\end{equation}

However, when we consider real scrambling matrices,
\eqref{eq:PeS} and \eqref{eq:PjPij} can be used again, except for the
following change in the expression of $P_j$, due to
the $t$-error correcting code with rate $<1$:
\begin{equation}
\begin{array}{rcl}
P_j & = & \sum_{i=t+1}^n{P_{i} \cdot P_{j|k}} \\
		& = & \sum_{i=t+1}^n{{n \choose i} P_0^i \left(1-P_0\right)^{n-i} \cdot 
		\frac{{k \choose j}{{n-k} \choose {i-j}}}{{n \choose i}}} \\
		& = & {k \choose j}\sum_{i=t+1}^n{{n-k \choose i-j}P_0^i\left(1-P_0\right)^{n-i}}.
\end{array}
\label{eq:PjReplaced}
\end{equation}
In \eqref{eq:PjReplaced}, $P_{i}$ is the probability that there are $i$ errors in a codeword and $P_{j|k}$
is the probability that exactly $j$ of such $i$ errors are within the $k$ information bits
associated with that codeword.

\begin{figure}[htb]
\begin{centering}
\includegraphics[width=83mm,keepaspectratio]{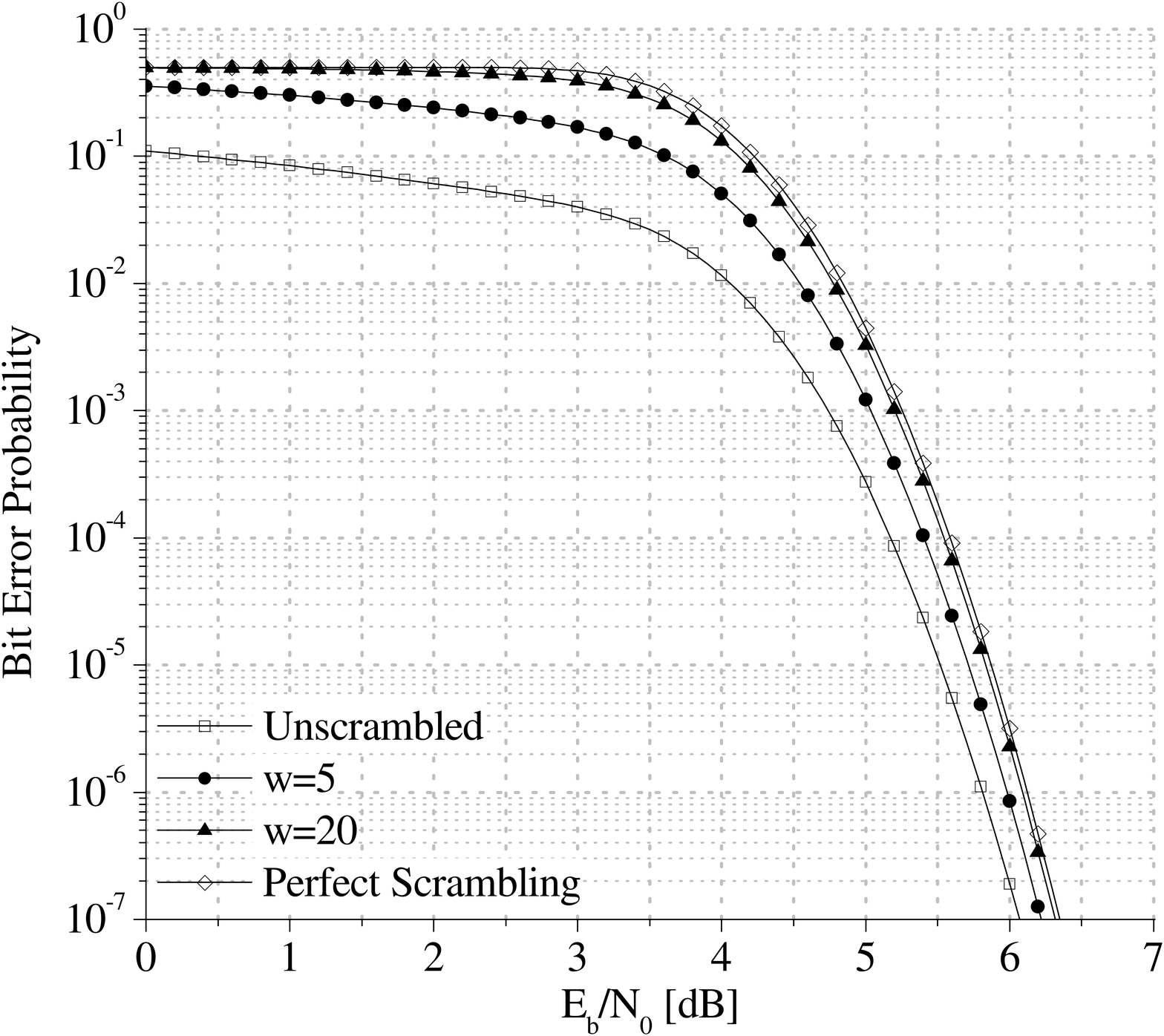}
\caption{Bit error probability for the $(511, 385)$ BCH code in the absence of scrambling,
with perfect scrambling, and for different values of the descrambling matrix column weight ($w$).
\label{fig:tErrorCorrecting}}
\par\end{centering}
\end{figure}

Some examples are shown in Fig. \ref{fig:tErrorCorrecting}, by
considering different levels of scrambling.
Through a comparison with Fig. \ref{fig:UnitaryRate}, we can observe that the code reduces the SNR for a given bit error probability (as expected) and, most important for our purposes, increases the slope of the $P_e^{S}$ and $P_e^{PS}$ curves.
From Fig. \ref{fig:tErrorCorrecting}, we can also note that, in this case, a descrambling matrix
with $w = 20$ (density $w/k \approx 0.05$) is sufficient to approach the effect of perfect scrambling.
Thus, when error correcting codes are used, reaching the performance of the perfect scrambler
requires even less dense descrambling matrices with respect to the case with unitary rate coding
(for which a density $w/k \geq 0.26$ was necessary; see Section \ref{subsec:UnitaryRate}).



\subsubsection{Scrambling of concatenated frames}
\label{subsubsec:tErrorCorrectingConcatenation}

Block-based scrambling can also be applied to transmissions adopting
$t$-error correcting codes with hard-decision decoding.

The effect of perfect scrambling on blocks of $L$ concatenated frames
can be estimated by using \eqref{eq:ConcPerfScrambUnitary} again, 
in which, obviously, the $P_f$ values must be updated by using \eqref{eq:tErrorCoding}.
The effect of real scrambling matrices working on $L$ concatenated
frames can instead be evaluated by resorting again to \eqref{eq:PeLS}.
In this case, $P_e^S$ can still be estimated by using \eqref{eq:PeS}
and \eqref{eq:PjPij}, and changing the expression of $P_j$ as in 
\eqref{eq:PjReplaced} to consider the $t$-error 
correcting code with rate $<1$.

\begin{figure}[htb]
\begin{centering}
\includegraphics[width=83mm,keepaspectratio]{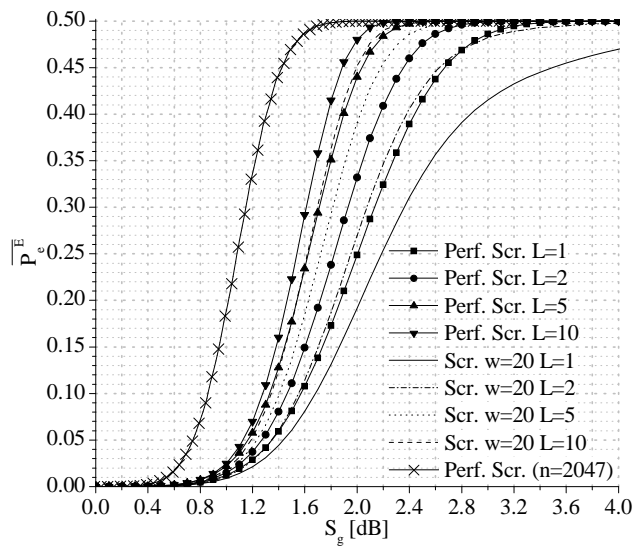}
\caption{Eve's bit error probability versus the security gap for $\overline{P_e^{\mathrm{B}}} = 10^{-5}$,
when using the $(511, 385)$ BCH code with several levels of concatenated scrambling ($L$),
under the hypothesis of perfect scrambling or with a descrambling matrix having column weight $w$.
The performance of a $(2047, 1541)$ BCH code is also reported (with perfect scrambling and no concatenation).
\label{fig:tErrorCorrecting_Conc}}
\par\end{centering}
\end{figure}

To evaluate the effect of concatenated scrambling on the security gap,
we again consider the $(511, 385)$ BCH code and fix $\overline{P_e^{\mathrm{B}}} = 10^{-5}$.
Fig. \ref{fig:tErrorCorrecting_Conc} reports the values of $\overline{P_e^{\mathrm{E}}}$, as a function 
of the security gap, for different values of $L$.
Also in this case, concatenated scrambling reduces the gap needed to increase $\overline{P_e^{\mathrm{E}}}$,
with respect to the case without concatenation ($L=1$).
We observed that the effect of a block descrambling matrix with row and column
weight $w \cdot L = 20 \cdot L$ approaches that of the perfect scrambler 
for increasing $L$.
Obviously, the convergence to perfect scrambling would also improve with the increasing $w$.

For the sake of comparison, Fig. \ref{fig:tErrorCorrecting_Conc} also includes
the curve of a $(2047, 1541)$ BCH code, capable of correcting $t=47$ errors.
This code has parameters that are about $4$ times those of the $(511, 385)$ BCH code, and
different from the case of unitary rate coding, the increased length provides better performance than
realizing concatenated scrambling (even with $L > 4$) over the shorter code.
Another example in this sense can be found in \cite{Baldi2010}.
This is expected, because adopting longer codes with higher error correction capability
increases the slope of the BER curves, thus reducing the security gap.
However, the use of longer codes also increases the decoding complexity to the point
that the system may become impractical.
On the contrary, the use of concatenated scrambling allows lowering the security gap,
while still using short codes with small decoding complexity.
In fact, increasing $L$ only affects the descrambling matrix size,
and, in addition, the descrambling complexity can be kept low by limiting 
its density.

\subsection{Nonsystematic LDPC Codes}
\label{sec:four}



\begin{figure}[htb]
\begin{centering}
\includegraphics[width=83mm,keepaspectratio]{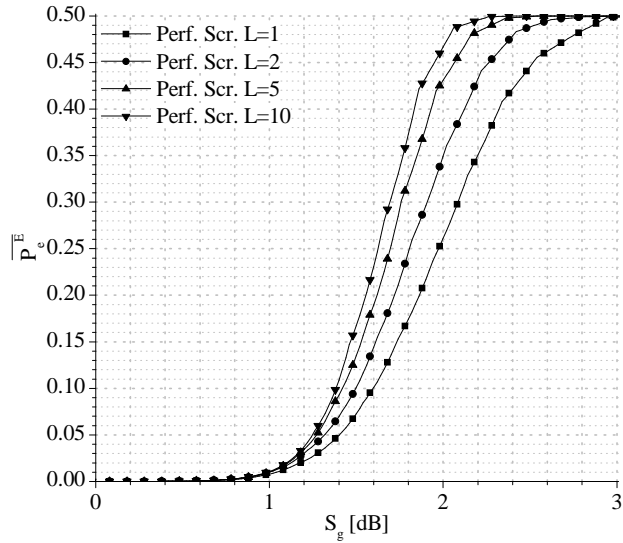}
\caption{Eve's bit error probability versus the security gap for $\overline{P_e^{\mathrm{B}}} = 10^{-5}$,
when using a $(511,385)$ LDPC code and different levels of concatenation ($L$),
under the hypothesis of perfect scrambling.
\label{fig:LDPC_Conc}}
\par\end{centering}
\end{figure}

For the sake of comparison, we consider an LDPC code having exactly the same parameters 
as the BCH code studied in the previous section, that is, $n=511$ and $k=385$. The secrecy rate is $R_s = R \approx 0.75$. The code has been designed through the progressive edge growth (PEG) algorithm 
\cite{Hu2001PEG} and has a lower triangular parity-check matrix.
This matrix and those of other LDPC codes considered
in our examples are available in \cite{Baldi2011website}.
Performance with perfect scrambling is shown in Fig. \ref{fig:LDPC_Conc} that reports the values of $\overline{P_e^{\mathrm{E}}}$ versus the security gap, for $\overline{P_e^{\mathrm{B}}} = 10^{-5}$ and different values of $L$.
It can be verified that by using real scrambling matrices
with a high density of ones in their inverse, the effect of block scrambling is 
almost coincident with that predicted through the perfect scrambler approximation. Both the case of scrambling on single frames ($L = 1$) and on concatenated frames ($L > 1$) are considered. Concatenated scrambling increases
the slope of the bit error probability curve, thus reducing the security gap.
Also, in this case, starting from the $P_f$ values for single frame scrambling (that are the same as for systematic transmission), the effect of a perfect scrambler on blocks 
of $L$ concatenated frames can be computed through \eqref{eq:ConcPerfScrambUnitary}.
However, contrary to the case of BCH codes, for a finite-length LDPC code, the values of $P_f$ must be estimated through numerical simulations.

\subsection{Comparison among the considered techniques}
\label{sec:five}

In this section, we compare the security gap performance achieved
by scrambled and punctured transmissions.
The approach based on puncturing does not benefit from the use of concatenation
(unless an interleaved code is used together with ARQ \cite{Harrison2010}).
Thus, for a fair comparison, we consider the case of single frame scrambling.

In \cite{Baldi2010}, a comparison of this kind has already been presented for the case of $R_s = 0.66$ and rather long codes.
The examples of BCH and LDPC codes discussed in the previous sections, instead, consider the case of $R_s = 0.75$ 
and rather short codes (i.e., with $k = 385$). 
To find a punctured code with the same parameters, we should start from a mother code with length $n' = 896$,
dimension $k' \geq k = 385$, and then puncture $k$ bits.
However, for these parameters and high rate, our simulations show that it is 
too difficult to find a punctured LDPC code with good performance, even at very high SNR,
and a lower secrecy rate becomes necessary.
Therefore, we fix a secrecy rate $R_s = 0.5$ for the puncturing-based system
and use an LDPC code having $n' = 1155$, $k' = k = 385$ for such setting, in which all the 
information bits are punctured (so that $R = R_s$).
In this way, the number of secret bits is fixed, though with different $R_s$.

\begin{figure}[htb]
\begin{centering}
\includegraphics[width=83mm,keepaspectratio]{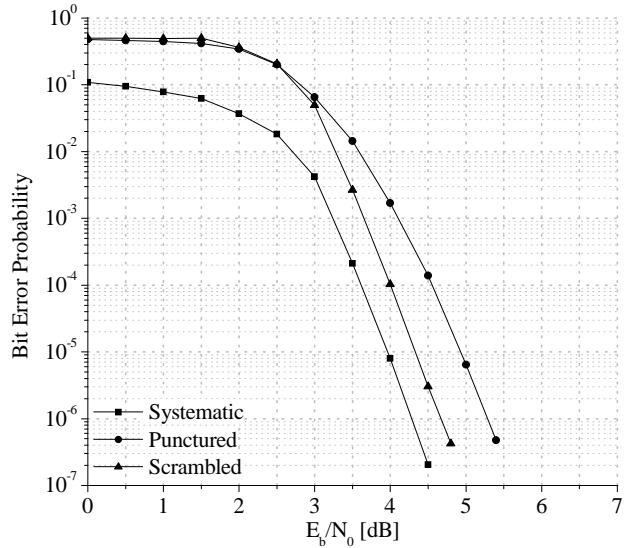}
\caption{Simulated bit error probability for a $(511, 385)$ LDPC code with systematic transmission
and perfect scrambling, in comparison with a punctured $(770, 385)$ LDPC code.
\label{fig:LDPC}}
\par\end{centering}
\end{figure}

Fig. \ref{fig:LDPC} shows the simulated performance.
We can observe that the systematic transmission achieves the best performance in
terms of error correction capability. However, it shows an important drawback concerning security, 
that is, a bit error probability $< 0.5$ even at low SNR.

The approach based on puncturing gives worse error correcting performance,
with a loss of about $1$ dB in the waterfall region with respect to systematic
LDPC coding (despite the lower rate of the punctured code).
However, the use of punctured bits for the secret message
provides higher bit error probability for low SNR.
Both such aspects benefit from the use of nonsystematic nonpunctured LDPC codes
based on scrambling: in this case, the performance loss with respect
to the systematic LDPC code is about $0.4$ dB in the waterfall region.

\begin{figure}[htb]
\begin{centering}
\includegraphics[width=83mm,keepaspectratio]{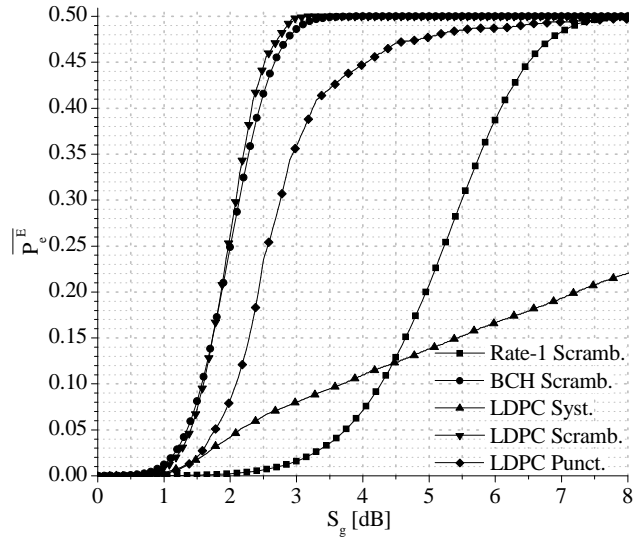}
\caption{Eve's bit error probability versus the secrecy gap
(at $\overline{\left. P_e \right|_\mathrm{B}} = 10^{-5}$) for
a ($k = n = 385$) unitary rate code, $(511, 385)$ BCH code, and $(511, 385)$ LDPC code
with perfect scrambling (without concatenation), in comparison with a punctured 
$(770, 385)$ LDPC code and a systematic $(511, 385)$ LDPC coded transmission.
\label{fig:BERvsGap}}
\par\end{centering}
\end{figure}

These facts also reflect on the security gap over the AWGN wire-tap channel.
As presented in the previous subsections, we fix $\overline{\left. P_e \right|_\mathrm{B}} = 10^{-5}$ 
and estimate $\overline{\left. P_e \right|_\mathrm{E}}$ as a function of the security gap $S_g$.
Fig. \ref{fig:BERvsGap} reports these curves for the considered transmission techniques.
We assume perfect scrambling for all techniques using scrambling.
The value of $k$ is fixed to $385$, while the secrecy rate is $0.75$ for
the scrambled codes and $0.5$ for the punctured code, as explained earlier.


As we can observe from the figure, the use of a systematic LDPC code gives a very slow
convergence of Eve's bit error probability to the ideal value of $0.5$.
Therefore, such technique requires a very high security gap for realistic values of $\overline{\left. P_e \right|_\mathrm{E}}$
(that are usually $\geq 0.4$).
The reason for such a slow convergence is systematic transmission: if we adopt
a nonsystematic unitary rate code, even after renouncing any error correction capability,
the performance is improved and $\overline{\left. P_e \right|_\mathrm{E}} \geq 0.4$ is reached for a gap value of $6.1$ dB.

These results are improved by nonsystematic transmission with error correcting codes.
If we implement nonsystematicity through puncturing, the condition $\overline{\left. P_e \right|_\mathrm{E}} \geq 0.4$
is achieved with a gap of $3.25$ dB.
By using scrambled transmission, both the BCH and LDPC code achieve very good performance.
The condition $\overline{\left. P_e \right|_\mathrm{E}} \geq 0.4$ is reached with a gap of $2.45$ and $2.33$ dB 
by the scrambled BCH and LDPC code, respectively, with a gain of about $0.8$ and $0.9$ dB
with respect to the approach based on puncturing.
Thus, the best performance is achieved by 
nonsystematic coded transmission through scrambling.
Such conclusion is even more evident if we consider that 
the performance of scrambled transmission can be further improved by concatenated scrambling,
as shown in the previous sections.

The same conclusion on the comparison with puncturing techniques can be drawn by using
codes with different parameters \cite{Baldi2010}; thus, it does not depend on the 
cases considered here.

\subsection{Complexity assessment for LDPC codes}
\label{subsec:Complexity}

As we have seen in the previous sections, nonsystematic transmission achieves better performance
in terms of security, with respect to systematic transmission.
In this section, we estimate the cost of such improvement in terms of complexity, for the case
of LDPC coded transmissions, and show that the advantage of scrambled transmission
over puncturing does not come at the cost of an increased complexity.

Let us denote the average number of binary operations for encoding as $C_{enc}$.
By using the generator matrix $\mathbf{G}$, which is generally dense,
a systematic code only requires computations for the redundancy part, that is,
$C_{enc} = k \cdot r /2$.
When scrambling is considered, the generator matrix becomes $\mathbf{G}' = \mathbf{S} \cdot \mathbf{G}$;
hence, $C_{enc} = k \cdot n / 2$.
In the case of puncturing, the generator matrix has a size $k \times n'$ and systematic form. 
However, puncturing coincides with eliminating the $k \times k$ identity block from the generator matrix;
hence, we again obtain $C_{enc} = k \cdot n / 2$.
When the codes have parity-check matrices in lower triangular form, similar to the PEG codes considered earlier,
some advantages result from performing encoding through back-substitution.
In this case, the number of binary operations approximately coincides with the total number of ones
in the matrix. 
This can be exploited by both puncturing and scrambling.
In the latter case, we can adopt sparse scrambling matrices, though with dense inverses,
and the encoding map carries out the following: i) computation
of $\mathbf{u}' = \mathbf{u} \cdot \mathbf{S}$ through multiplication of $\mathbf{u}$ by the 
sparse matrix $\mathbf{S}$ and ii) LDPC encoding of $\mathbf{u}'$ through back-substitution.

With regard to the decoding complexity, we consider the implementation of SPA proposed in \cite{Hu2001}, 
according to which the number of binary operations per decoded codeword is
$C_{SPA}=I_{ave}\cdot n\left[q\left(8d_{v}+12 k/n -11\right)+d_{v}\right]$.
In this expression, $I_{ave}$ is the average number of decoding iterations,
$q$ is the number of quantization bits,
and $d_{v}$ is the average degree of variable nodes in 
the code Tanner graph.
When the code is punctured, its length is $n' > n$, because the 
decoder works on the full code to recover the values of the punctured bits.
Finally, when scrambling is used, we must consider additional $k \cdot w$ operations
to obtain the decoding complexity ($C_{dec}$), due to multiplication of the 
decoded information word by the inverse scrambling matrix.

As we have verified through simulations, the SPA decoder working on
punctured codes requires a higher average number of iterations to converge with respect to the
cases without puncturing.
This partially counterbalances the operations necessary for descrambling.

\begin{table}[htb]
\renewcommand{\arraystretch}{0.9}
\caption{Complexity assessment for systematic, scrambled, and punctured LDPC codes at BER $\approx 10^{-5}$.}
\label{tab:Complexity}
\centering
\begin{tabular}{|c|c|c|c|}
\hline
				& Systematic LDPC & Scrambled LDPC & Punctured LDPC \\
\hline
\hline
$n$ $(n')$	& $511$		& $511$		& $1155$ \\
\hline
$k$				& $385$		& $385$		& $385$ \\
\hline
$z$				& -				& -				& $385$ \\
\hline
$R$				& $0.75$	& $0.75$	& $0.5$ \\
\hline
$d_v$			& $3.8$		& $3.8$		& $3.5$ \\
\hline
$w$				& - 			& $193$		& - \\
\hline
$E_b/N_0$	& $4.0$		& $4.3$		& $4.9$ \\
\hline
$I_{ave}$	& $2$			& $2$			& $9.2$ \\
\hline
$q$				& $8$			& $8$			& $8$ \\
\hline
$C_{enc}$	& $2^{14.6}$ $(2^{10.9})$ & $2^{16.6}$ $(2^{12.2})$ & $2^{17.2}$ $(2^{12.0})$ \\
\hline
$C_{dec}$	& $2^{17.9}$	& $2^{18.2}$	& $2^{20.8}$ \\
\hline
\end{tabular}
\end{table}

The complexity estimates for the systems adopting systematic, scrambled,
and punctured LDPC coded transmission, for a target BER $\approx 10^{-5}$, are 
reported in Table \ref{tab:Complexity}, where
$z$ is the number of punctured bits.
The table confirms that systematic transmission has the lowest complexity,
although it has the worst performance under the security viewpoint.
The systems using scrambling
and puncturing have comparable encoding and decoding complexity.
When the LDPC matrix is in lower triangular form, all systems can 
reduce the number of encoding operations by exploiting back-substitution
and, in the case of scrambled codes, sparse scrambling matrices as well.
The corresponding $C_{enc}$ values are reported between brackets in the table.
For the system with scrambling, the
encoding complexity value considers a scrambling matrix $\mathbf{S}$ with row and column weight equal to $7$.


\section{Coded transmission with ARQ}
\label{sec:arq}

As shown in the previous sections, coded transmission based on scrambling is able to
significantly reduce the security gap.
However, according to the previous analysis, a prefixed level of physical layer security seems achievable only when Bob has 
a better channel than Eve.
On the other hand, it is known that when Bob's and Eve's channels have the same quality,
or Eve's channel is better than Bob's, a feedback mechanism is needed
to achieve physical layer security \cite{Leung-Yan-Cheong1978, Lai2008}.

We investigate this case by considering a very simple feedback mechanism, based
on integrity checks and ARQ. The latter was already used
in the implementation of physical layer security schemes, although in different scenarios 
\cite{Tang2009, Omar2009, Harrison2010}.
Obviously, the request for retransmission must be allowed only for Bob; hence,
some form of authentication between Alice and Bob is required.
However, retransmitted packets are also available to Eve, through her channel.


In the present study, we have used ARQ together with forward error correcting codes, that is,
a HARQ scheme.
Several implementations are possible for HARQ:
the two main families are those using incremental redundancy \cite{Mandelbaum1974}
and soft combining \cite{Zepernick2002, Holland2005}.
When soft-decision MAP decoding algorithms are adopted, 
a common approach consists of using the reliability values obtained after each (failed) decoding attempt
as \textit{a priori} values for decoding the next transmission of the same frame \cite{Zepernick2002}.
The soft-combining strategy proposed in \cite{Holland2005}, capable of improving the performance
of the approach in \cite{Zepernick2002}, coincides with averaging the channel outputs
after multiple transmissions of the same frame.
Such scheme is a valuable benchmark, because it provides good performance and
can be implemented with any family of codes (exploiting both hard- and soft-decision decoding).


In the HARQ protocol that we considered, Bob can exploit a number of transmissions $Q \leq Q_{\max}$ for decoding each frame
and Eve receives all retransmissions requested by Bob.
It should be noted that the integrity check mechanism, based on parity-checks, is exposed to
undetected errors, that is, transitions of the received codeword to near codewords.
In such case, the integrity of the frame is erroneously verified. However, for the values of code parameters that are of interest in practical applications, undetected errors are rare and therefore can be neglected.

The solution that we have adopted consists of trying decoding on the average of the channel outputs
for all the received transmissions of each frame.
In this case, when decoding is performed over $Q \leq Q_{\max}$ transmissions of a frame, we
can estimate the frame error probability as:
\begin{equation}
P_f^{(Q)}\left(\frac{E_b}{N_0}\right) \approx P_f \left(\frac{E_b}{N_0} \cdot Q\right),
\label{eq:PfQ}
\end{equation}
where $P_f\left(E_b/N_0\right)$ is the frame error probability in the absence of ARQ.
Applying \eqref{eq:PfQ} coincides with dividing the noise variance by $Q$
due to averaging of the channel outputs.
Approximation results from the observation that \eqref{eq:PfQ} neglects the correlation existing between subsequent retransmissions (a retransmission is requested only when the previous transmission fails).

To verify the impact of this approximation, we can consider the example
of a \textit{t}-error correcting code with rate $R < 1$, used in an ARQ scheme with $Q_{\max} = 2$ .
Let us define $g(x) = \frac{e^{-x^2R/N_0}}{\sqrt{\pi N_0/R}}$
and the following probabilities:
$P_1 = \int_{-\infty}^{\alpha}{g(x) \int_{\beta}^{\infty}{g(y)dy} dx}$, 
$P_2 = \int_{\alpha}^{\infty}{g(x) \int_{\beta}^{\infty}{g(y)dy} dx}$, 
$P_3 = \int_{\alpha}^{\infty}{g(x) \int_{-\infty}^{\beta}{g(y)dy} dx}$, 
$P_4 = \int_{-\infty}^{\alpha}{g(x) \int_{-\infty}^{\beta}{g(y)dy} dx}$,
where
$\alpha = -\sqrt{E_b}$ and $\beta = -2\sqrt{E_b}-x$.
It can be proved that for this case, the frame error probability
at the second decoding attempt, conditioned on the fact that the first transmission failed (i.e., taking into account the correlation between subsequent transmissions), can be expressed as:
\begin{eqnarray}
P_f^{(2)}& = 	& 1 - {P_f}^{-1} \cdot \sum_{i=t+1}^n {n \choose i} \sum_{j=i-t}^i {i \choose j} P_1^j P_4^{i-j} \nonumber \\
				 &		& \sum_{l=0}^{t+j-i} {n-i \choose l} P_3^l P_2^{n-i-l},
\label{eq:PfQnonGauss}
\end{eqnarray}
where $P_f$ is given by \eqref{eq:tErrorCoding}.
The values of the frame error probability provided by \eqref{eq:PfQnonGauss} for the $(511, 385)$ BCH code, capable of correcting $t = 14$ errors, are very close to those resulting from \eqref{eq:PfQ}, for $P_f \geq 10^{-8}$. The analysis could be extended to higher values of $Q_{\max}$, yielding increasingly complicated expressions.
However, even for these more general cases, we have verified through simulations that the impact of the correlation is always negligible, at least for our choice of the parameters. Thus, \eqref{eq:PfQ} can be used as a good approximation of the real behavior.

In the system model that we have adopted, Bob is always able to request retransmission of a frame, when
needed, that is, after a decoding failure, up to $Q_{\max}$. Hence, the probability that he receives $Q \geq 1$ transmissions 
of a frame coincides with his frame error probability after $Q-1$ transmissions, that is:
\begin{equation}
\left. P_R^{(Q)} \right|_{\mathrm{B}} = \prod_{i=0}^{Q-1} P_f^{(i)}\left(\left. \frac{E_b}{N_0} \right|_\mathrm{B}\right),
\label{eq:PRQB}
\end{equation}
with $P_f^{(0)} = 1$.
On the contrary, Eve can benefit from the retransmission of an erred frame only when Bob fails to decode the same frame.
Hence, the probability that she receives $Q \geq 1$ (useful) transmissions of a frame is:
\begin{equation}
\left. P_R^{(Q)} \right|_{\mathrm{E}} = \prod_{i=0}^{Q-1} P_f^{(i)}\left(\left. \frac{E_b}{N_0} \right|_\mathrm{B}\right) P_f^{(i)}\left(\left. \frac{E_b}{N_0} \right|_\mathrm{E}\right).
\label{eq:PRQE}
\end{equation}
Finally, Bob's and Eve's frame error probability is:
\begin{equation}
\left. P_f^\mathrm{ARQ}\right|_{\mathrm{B}/\mathrm{E}} = 1 - \displaystyle\sum_{i=1}^{Q_{\max}} \left. P_R^{(i)} \right|_{\mathrm{B}/\mathrm{E}} \cdot \left[1 - P_f^{(i)}\left(\left. \frac{E_b}{N_0} \right|_{\mathrm{B}/\mathrm{E}}\right) \right].
\label{eq:PfARQ}
\end{equation}

From now on, we assume $\left. \frac{E_b}{N_0} \right|_\mathrm{B} \leq \left. \frac{E_b}{N_0} \right|_\mathrm{E}$,
which implies to have $S_g \leq 1$, to show that the adoption of HARQ permits to achieve a sufficient level of
physical layer security even when Eve's channel is better than Bob's.

Next, we provide some results on the use of the HARQ scheme
with the $t$-error correcting codes and LDPC codes considered in the previous
sections.
The case of unitary rate codes cannot be considered, because
it does not allow any integrity check mechanism.

\subsection{ARQ with $t$-Error Correcting codes}
\label{subsec:arq-bch}

\begin{figure}[htb]
\begin{centering}
\includegraphics[width=83mm,keepaspectratio]{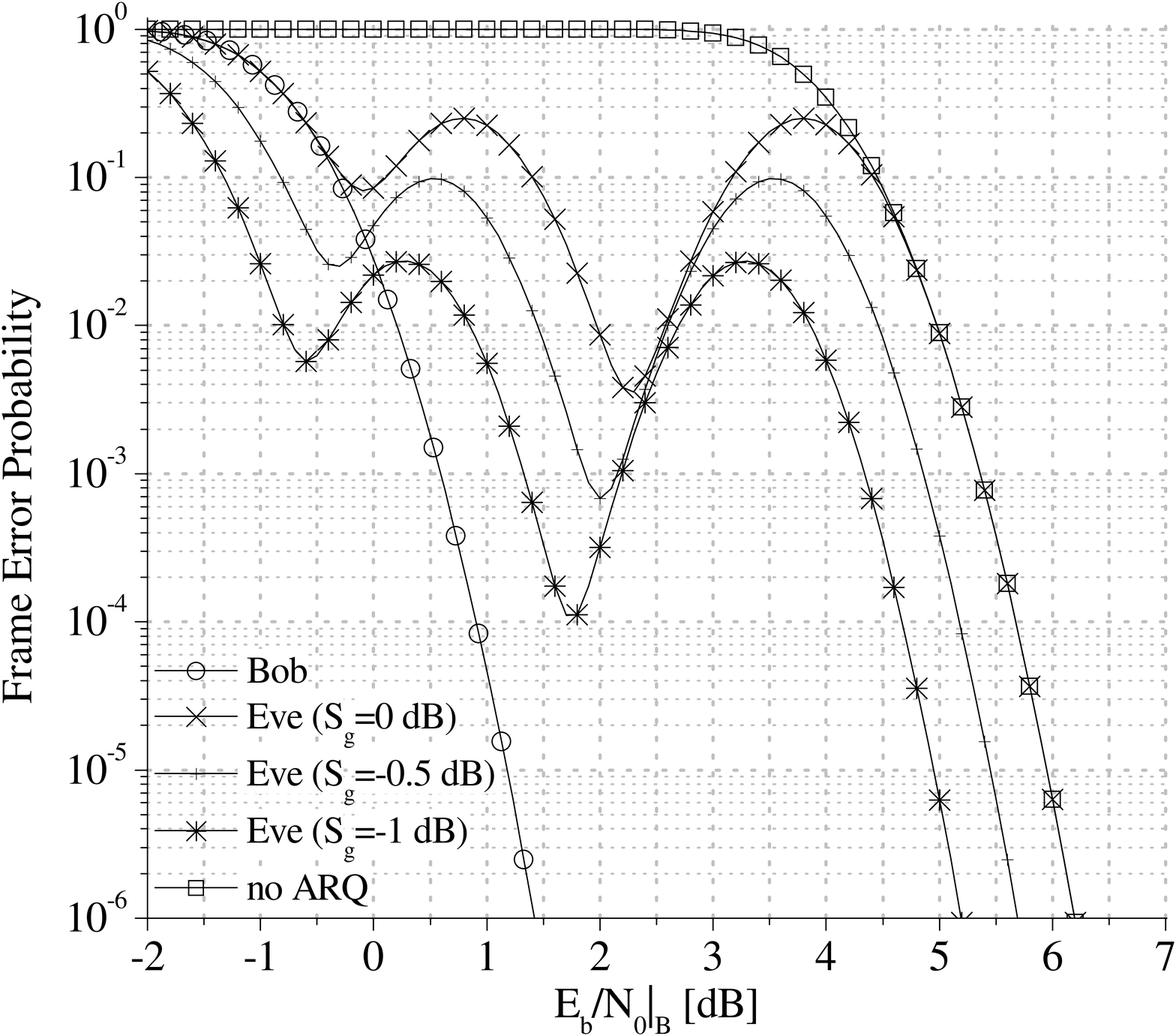}
\caption{Frame error probability versus Bob's SNR for the $(511, 385)$ BCH code with soft-combining HARQ ($Q_{\max} = 3$)
and different values of security gap ($S_g$).
\label{fig:ARQ-BCH}}
\par\end{centering}
\end{figure}

Fig. \ref{fig:ARQ-BCH} shows the frame error probability curves for the $(511, 385)$ BCH code, obtained
through \eqref{eq:PfARQ} and for different values of $S_g$. 
The maximum number of transmissions is $Q_{\max} = 3$.
We have focused on the frame error probability because it also models perfect scrambling.
When ARQ is not used, Eve's performance only depends on her channel SNR, and the same occurs for Bob.
Thus, the curve without ARQ applies both to Bob and Eve, as in the previous sections.

However, when ARQ is used, Eve's performance also depends on Bob's SNR, as explained in the following.
We can observe from the figure that Bob's frame error probability is monotonically decreasing,
while Eve's curves exhibit an oscillating behavior.
This is because performance is determined by the values of $P_f^{(1)}\left(\left. \frac{E_b}{N_0} \right|_\mathrm{B}\right)$;
that is, Bob's frame error probability without ARQ. For simplicity, such probability will be denoted as $P_f^{(B)}$ in the following.
For very low SNR, $P_f^{(B)} = 1$ and Bob asks for $Q_{\max}$ transmissions of all frames;
Eve also benefits by this fact, and her performance, for $S_g = 0$ dB, is coincident with that of Bob.
Both of them show a gain of $\log_{10}(Q_{\max}) = 4.77$ dB against that without ARQ.
For $S_g < 0$ dB, Eve's gain is greater than Bob's, as expected.
When the SNR increases and $P_f^{(B)}$ becomes smaller than (though remaining close to) $1$, only a fraction of the frames is transmitted $Q_{\max}$ times, and Eve misses most of the retransmissions that she would need for a correct decoding. Correspondingly, Eve's performance deteriorates and her error probability curve, after having reached a minimum (dependent on the value of $S_g$), rises until it catches up with the ARQ curve for a maximum of two retransmissions.
A further decrease in $P_f^{(B)}$ produces a similar behavior: when the average number of retransmissions (requested by Bob) approaches $1$, Eve's error probability curve reaches a second minimum and then rises again approaching the curve without ARQ (which coincides with $P_f^{(B)}$ in the case $S_g = 0$ dB, while it shows a gain equal to $S_g$ in the case of Eve's channel better than Bob's).

Such characteristic behavior would appear for any choice of $Q_{\max}$, with an alternation of minima and maxima that can be usefully exploited for the security issue. In this perspective, however, we observe that in the region where Bob achieves
low frame error probabilities, Eve's frame error probability is always too low to guarantee 
a sufficient level of security.

The way to restore the security condition is to use concatenated scrambling. This can be verified through numerical examples. Let us consider to use a concatenated scrambler with $L = 170$.
By applying \eqref{eq:ConcPerfScrambUnitary}, the region of $\left. P_f^{{\rm ARQ}}\right|_\mathrm{E} \geq 10^{-2}$ translates into 
$\left. P_f^{{\rm ARQ}}\right|_\mathrm{E} > 0.82$ and, for any of the considered values of $S_g$, there is a rather large range of values
of $E_b/N_0 > 2.5$ dB, where this occurs.
In the same region, Bob's frame error probability is $\left. P_f^{{\rm ARQ}}\right|_\mathrm{B} < 10^{-7}$, and it becomes
$\left. P_f^{{\rm ARQ}}\right|_\mathrm{B} < 1.7 \cdot 10^{-5}$ after the application of the concatenated scrambler.
Thus, under perfect scrambling, we have $P_e^{\mathrm{E}} > 0.4$
and $P_e^{\mathrm{B}} < 1 \cdot 10^{-5}$, which restore security.

With respect to the examples in the previous section, where concatenation was
used to reduce the security gap, here, a higher value of $L$ is required.
This is because, in this context, the security gap is very small (down to $-1$ dB),
and hence, a higher level of concatenation is necessary.
This influences latency, but, for very short codes as those considered,
concatenating a number of frames on the order of $170$ should still have a tolerable effect.

\subsection{ARQ with LDPC codes}
\label{subsec:arq-ldpc}

\begin{figure}[htb]
\begin{centering}
\includegraphics[width=83mm,keepaspectratio]{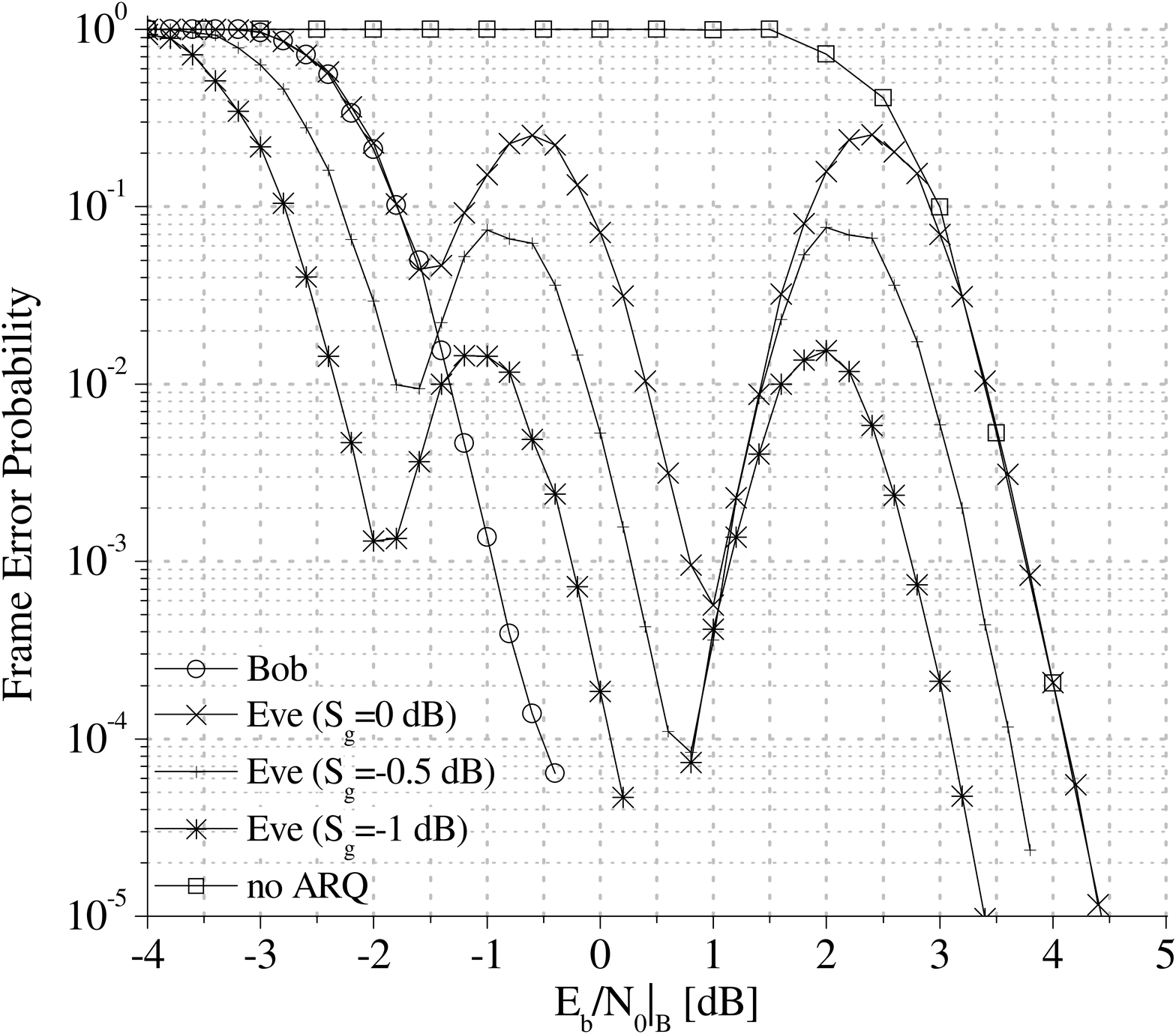}
\caption{Frame error probability versus Bob's SNR for the $(511, 385)$ LDPC code with soft-combining HARQ ($Q_{\max} = 3$)
and different values of security gap ($S_g$).
\label{fig:ARQ-LDPC}}
\par\end{centering}
\end{figure}

A situation similar to that presented in Section \ref{subsec:arq-bch} is observed for ARQ with LDPC codes.
Examples of frame error probabilities, estimated through numerical simulations, are reported in Fig. \ref{fig:ARQ-LDPC}.
From the figure, it can be noticed that Bob's and Eve's error probability curves have the same
behavior discussed for hard-decision decoded BCH codes.
Also in this case, we could observe the existence of a region (for $E_b/N_0 > 1.5$ dB) where Eve's frame
error probability is $\geq 10^{-2}$, while that of Bob's (according to the trend of the simulated curve) 
will become $< 10^{-7}$. Hence, a concatenated scrambler with $L = 170$ still 
allows to achieve the desired level of physical layer security.

Thus, by adopting LDPC codes as well, HARQ allows achieving a prefixed security 
level even when Eve's channel is not worse than that of Bob.
Furthermore, the use of LDPC codes permits us to reduce the SNR working point by more than $1$ dB, with respect to
hard-decision decoded BCH codes.
Such advantage would be even more evident by adopting longer codes.

\section{Secrecy performance of the considered codes}
\label{sec:Secrecy}

In a recent paper \cite{WongWong2011a}, Wong et al. used the equivocation rate about the message at the wire-tapper
to measure the secrecy performance of a coding scheme based on punctured LDPC codes
with BPSK modulation.
A capacity-equivocation region can be defined, which contains all the achievable rate-equivocation pairs ($R_s, R_e$).
Under suitable assumptions, the expression of $R_e$ is given by:
\begin{equation}
R_e = R - C\left(\sqrt{\left. \frac{E_b}{N_0} \right|_\mathrm{E}}\right),
\label{eq:Wong}
\end{equation}
where $C(t)$ is the channel capacity of the AWGN channel with BPSK input. 
The value of $R_e$ is upper limited by the secrecy capacity $C_s$, and $R_e \leq R_s \leq C\left(\sqrt{\left. \frac{E_b}{N_0} \right|_\mathrm{B}}\right)$.

Instead of (\ref{eq:Wong}), a normalized value ${\tilde R}_e = \frac{R_e}{R_s}$, called the ``fractional equivocation,'' may be preferred. As we consider $R = R_s$, maximizing ${\tilde R}_e$ for a given value of $R_s$ and a fixed security gap consists of finding the minimum $\left. \frac{E_b}{N_0} \right|_\mathrm{E}$ such that $\left. P_e \right|_\mathrm{B} \leq \overline{\left. P_e \right|_\mathrm{B}}$ and $0.5 \geq \left. P_e \right|_\mathrm{E} \geq \overline{\left. P_e \right|_\mathrm{E}}$. Accordingly, $C\left(\sqrt{\left. \frac{E_b}{N_0} \right|_\mathrm{E}}\right)$ is also minimum. It is evident that this implies a trade-off. Therefore, for a fixed value of $S_g > 1$, we have considered a number of scrambled BCH and LDPC codes that satisfy the constraints on both the thresholds and compute the value of ${\tilde R}_e$. Some examples are shown in Table \ref{tab:Equivocation}, where we have set $R_s = 0.43$ and $S_g = 4.4$ dB. These values are equal to those considered in \cite{WongWong2011a}, thus permitting a comparison with the coding scheme proposed in that paper.
Two of the LDPC codes in Table \ref{tab:Equivocation} are almost regular, with column weight $\approx 3$.
In addition, an irregular LDPC code is considered, designed using the PEG algorithm available in \cite{MacKay2011}
and a degree distribution with concentrated check node degrees and maximum variable node degree equal to $11$,
optimized through density evolution.
Their performance is estimated by considering perfect scrambling. Any scrambling matrix capable of reaching this condition can be equivalently applied. As shown in Section \ref{sec:three}, matrices of this kind are easy to design.

\begin{table} [htb]
\caption{Fractional equivocation rate for some scrambled BCH and LDPC codes by assuming $R_s = R = 0.43$ and $S_g = 4.4$ {\normalfont dB}.}
\label{tab:Equivocation}
\centering
\begin{tabular}{ccccc}
\hline
Code & $n$ & $k$ & $\left. E_b/N_0 \right|_\mathrm{B}$@$10^{-5}$ [dB] & ${\tilde R}_e$ \tabularnewline
\hline
BCH & $511$ & $220$ & $5.63$ & $0.2496$\tabularnewline
BCH & $2047$ & $881$ & $5.48$ & $0.2697$\tabularnewline
reg. LDPC& $2000$ & $860$ & $2.25$ & $0.6101$\tabularnewline
irr. LDPC& $2000$ & $860$ & $1.8$ & $0.6446$\tabularnewline
reg. LDPC & $50000$ & $21500$ & $1.1$ & $0.6928$\tabularnewline
\hline
\end{tabular}
\end{table}

The reliability is assumed to be equal to $10^{-5}$, while the security threshold is $0.49$, as in \cite{WongWong2011a}. 
Table \ref{tab:Equivocation} shows that by requiring greater SNR for Bob's channel, when compared with the LDPC codes, the
solutions based on BCH codes exhibit significantly smaller equivocation rates.
The value of ${\tilde R}_e$ increases, as expected, for increasing code lengths. 
In the case of regular LDPC codes of the type considered, the limit value (for $n \rightarrow \infty$) can be estimated by using density evolution \cite{Chung2001}, and results in ${\tilde R}_e = 0.7005$. 
This value compares favorably with that reported in \cite{WongWong2011a}, where ${\tilde R}_e = 0.7$ is obtained for a punctured irregular LDPC code.
Moreover, the results provided by the density evolution analysis can be referred to the use of a bitwise MAP decoder,
which is an optimal choice for Eve.

The value of ${\tilde R}_e$ achievable by the code proposed in \cite{Klinc2009a} is ${\tilde R}_e = 0.68$ \cite{WongWong2011a}. Thus, we can conclude that scrambled and almost regular LDPC codes permit us to achieve secrecy performance that is similar to and even better than that obtained by optimized punctured irregular LDPC codes. By considering scrambled and irregular LDPC codes, further improvements can be achieved, because the irregular LDPC code with $n=2000$ reaches ${\tilde R}_e = 0.6446$,
and the limit value resulting from density evolution for $R_s = 0.43$ is ${\tilde R}_e = 0.7567$, which is very close to the boundary of the capacity-(fractional) equivocation region. The latter, for $R_s = 0.43$, results in $C_s/R_s = 0.7863$.

\section{Conclusion}
\label{sec:six}
In this study, we have investigated the use of codes with scrambling, concatenation, and ARQ for physical layer security
over the AWGN wire-tap channel.
We have provided some theoretical tools that help estimating the bit and frame error probabilities
of scrambled transmissions.
Based on these tools and numerical simulations, we have studied some examples of codes to assess 
the security performance of practical transmission schemes.

Our results show that nonsystematic transmission is capable of reducing the security gap in terms of
SNR that is needed between Bob's and Eve's AWGN channels to achieve physical layer security.
We have compared nonsystematic transmission implemented through scrambling and puncturing,
and showed that the former is able to outperform the latter, requiring a smaller security gap.
The use of concatenated scrambling can further reduce the security gap. Moreover, it becomes mandatory 
when Eve's channel is not worse than Bob's, and HARQ protocols are used to achieve security and reliability.

Finally, we have shown that the proposed approach also has a good performance with regard to the maximization 
of the equivocation rate under the BPSK constraint, as it gives values very close to the ultimate capacity bounds.

\newcommand{\BIBdecl}{\setlength{\itemsep}{0.005\baselineskip}}


\begin{thebibliography}{10}
\providecommand{\url}[1]{#1}
\csname url@samestyle\endcsname
\providecommand{\newblock}{\relax}
\providecommand{\bibinfo}[2]{#2}
\providecommand{\BIBentrySTDinterwordspacing}{\spaceskip=0pt\relax}
\providecommand{\BIBentryALTinterwordstretchfactor}{4}
\providecommand{\BIBentryALTinterwordspacing}{\spaceskip=\fontdimen2\font plus
\BIBentryALTinterwordstretchfactor\fontdimen3\font minus
  \fontdimen4\font\relax}
\providecommand{\BIBforeignlanguage}[2]{{%
\expandafter\ifx\csname l@#1\endcsname\relax
\typeout{** WARNING: IEEEtran.bst: No hyphenation pattern has been}%
\typeout{** loaded for the language `#1'. Using the pattern for}%
\typeout{** the default language instead.}%
\else
\language=\csname l@#1\endcsname
\fi
#2}}
\providecommand{\BIBdecl}{\relax}
\BIBdecl

\bibitem{Wyner1975}
A.~D. Wyner, ``The wire-tap channel,'' \emph{Bell Syst. Tech. J.}, vol.~54,
  no.~8, pp. 1355--1387, Oct. 1975.

\bibitem{Klinc2011}
D.~Klinc, J.~Ha, S.~McLaughlin, J.~Barros, and B.-J.~Kwak, ``{LDPC} codes for
  the {Gaussian} wiretap channel,'' \emph{{IEEE} Trans. Inf. Forensics Security},
  vol. 6, no. 3, pp. 532--540, Sept. 2011. 

\bibitem{Klinc2009a}
------, ``{LDPC} codes for the {Gaussian} wiretap channel,'' in \emph{Proc.
  {IEEE} Information Theory Workshop ({ITW 2009})}, Taormina, Italy, Oct. 2009,
  pp. 95--99.

\bibitem{Klinc2009}
------, ``{LDPC} codes for physical layer security,'' in \emph{Proc. {IEEE}
  Global Telecommunications Conference ({GLOBECOM 2009})}, Honolulu, HI, Nov.
  2009, pp. 1--6.

\bibitem{WongWong2011a}
C.~W. Wong, T.~F. Wong, and J.~M. Shea, ``{LDPC} code design for the
  {BPSK}-constrained {Gaussian} wiretap channel,'' in \emph{Proc. {IEEE} 
  {GLOBECOM} Workshops 2011}, Houston, TX, Dec. 2011, pp. 898--902.   

\bibitem{WongWong2011}
------, ``Secret-sharing {LDPC} codes for the {BPSK}-constrained {Gaussian}
  wiretap channel,'' \emph{{IEEE} Trans. Inf. Forensics Security}, vol. 6, 
  no. 3, pp. 551--564, Sep. 2011.  

\bibitem{Tang2009}
X.~Tang, R.~Liu, P.~Spasojevic, and H.~V. Poor, ``On the throughput of secure
  hybrid-{ARQ} protocols for {Gaussian} block-fading channels,'' \emph{{IEEE}
  Trans. Inform. Theory}, vol.~55, no.~4, pp. 1575--1591, Apr. 2009.

\bibitem{Baldi2010}
M.~Baldi, M.~Bianchi, and F.~Chiaraluce, ``Non-systematic codes for physical
  layer security,'' in \emph{Proc. {IEEE } Information Theory Workshop (ITW
  2010)}, Dublin, Ireland, Aug. 2010.

\bibitem{Baldi2011}
------, ``Increasing physical layer security through scrambled codes and
  {ARQ},'' in \emph{Proc. {IEEE} International Conference on Communications
  ({ICC 2011})}, Kyoto, Japan, Jun. 2011.

\bibitem{Richardson2001}
T.~J. Richardson and R.~L. Urbanke, ``The capacity of low-density parity-check
  codes under message-passing decoding,'' \emph{{IEEE} Trans. Inform. Theory},
  vol.~47, no.~2, pp. 599--618, Feb. 2001.

\bibitem{Massey1969}
J.~L. Massey, ``Shift-register synthesis and {BCH} decoding,'' \emph{{IEEE}
  Trans. Inform. Theory}, vol.~15, no.~1, pp. 122--127, Jan. 1969.

\bibitem{Hagenauer1996}
J.~Hagenauer, E.~Offer, and L.~Papke, ``Iterative decoding of binary block and
  convolutional codes,'' \emph{{IEEE} Trans. Inform. Theory}, vol.~42, no.~2,
  pp. 429--445, Mar. 1996.

\bibitem{Heys1995}
H.~Heys and S.~Tavares, ``Avalanche characteristics of substitution-permutation
  encryption networks,'' \emph{{IEEE} Trans. Comput.}, vol.~44, no.~9, pp.
  1131--1139, Sep. 1995.

\bibitem{Torrieri1984}
D.~Torrieri, ``The information-bit error rate for block codes,'' \emph{{IEEE}
  Trans. Commun.}, vol.~32, no.~4, pp. 474--476, Apr. 1984.

\bibitem{Hu2001PEG}
X.~Y. Hu and E.~Eleftheriou, ``Progressive edge-growth {Tanner} graphs,'' in
  \emph{Proc. {IEEE} Global Telecommunications Conference ({GLOBECOM}'01)}, San
  Antonio, Texas, Nov. 2001, pp. 995--1001.

\bibitem{Baldi2011website}
\BIBentryALTinterwordspacing
M.~Baldi. (2011) Website. [Online]. Available:
  \url{http://sites.google.com/site/marcobaldi/publications/tifs-2011-attachments}
\BIBentrySTDinterwordspacing

\bibitem{Hu2001}
X.~Y. Hu, E.~Eleftheriou, D.~M. Arnold, and A.~Dholakia, ``Efficient
  implementations of the sum-product algorithm for decoding {LDPC} codes,'' in
  \emph{Proc. {IEEE} Global Telecommunications Conference ({GLOBECOM '01})},
  vol.~2, San Antonio, TX, Nov. 2001, pp. 1036--1036E.

\bibitem{Leung-Yan-Cheong1978}
S.~Leung-Yan-Cheong and M.~Hellman, ``The {Gaussian} wire-tap channel,''
  \emph{{IEEE} Trans. Inform. Theory}, vol.~24, no.~4, pp. 451--456, Jul. 1978.

\bibitem{Lai2008}
L.~Lai, H.~El~Gamal, and H.~V. Poor, ``The wiretap channel with feedback:
  Encryption over the channel,'' \emph{{IEEE} Trans. Inform. Theory}, vol.~54,
  no.~11, pp. 5059--5067, Nov. 2008.

\bibitem{Omar2009}
Y.~Omar, M.~Youssef, and H.~El~Gamal, ``{ARQ} secrecy: From theory to
  practice,'' in \emph{Proc. {IEEE } Information Theory Workshop (ITW 2009)},
  Taormina, Italy, Oct. 2009, pp. 6--10.

\bibitem{Harrison2010}
W.~K. Harrison, J.~Almeida, D.~Klinc, S.~W. McLaughlin, and J.~Barros,
  ``Stopping sets for physical-layer security,'' in \emph{Proc. {IEEE }
  Information Theory Workshop (ITW 2010)}, Dublin, Ireland, Aug. 2010.

\bibitem{Mandelbaum1974}
D.~Mandelbaum, ``An adaptive-feedback coding scheme using incremental
  redundancy,'' \emph{{IEEE} Trans. Inform. Theory}, vol.~20, no.~3, pp.
  388--389, May 1974.

\bibitem{Zepernick2002}
H.-J. Zepernick, B.~Rohani, and A.~Caldera, ``Soft-combining technique for
  {LUEP} codes,'' \emph{Electron. Lett.}, vol.~38, no.~5, pp. 234--235, Feb.
  2002.

\bibitem{Holland2005}
I.~Holland, H.-J. Zepernick, and M.~Caldera, ``Soft combining for hybrid
  {ARQ},'' \emph{Electron. Lett.}, vol.~41, no.~22, pp. 1230--1231, Oct. 2005.

\bibitem{MacKay2011}
\BIBentryALTinterwordspacing
D.~MacKay. (2011) Source code for {Progressive Edge Growth} parity-check matrix
  construction. [Online]. Available:
  \url{http://www.inference.phy.cam.ac.uk/mackay/PEG_ECC.html}
\BIBentrySTDinterwordspacing

\bibitem{Chung2001}
S.-Y. Chung, T.~J. Richardson, and R.~L. Urbanke, ``Analysis of sum-product
  decoding of low-density parity-check codes using a {Gaussian}
  approximation,'' \emph{{IEEE} Trans. Inform. Theory}, vol.~47, no.~2, pp.
  657--670, Feb. 2001.

\end{thebibliography}
\end{document}